\documentclass[10pt]{iopart}

\usepackage{iopams}  
\usepackage{epsfig}

\begin{document}

\title{Quantum two and three person duels}

\author{Adrian P. Flitney
and Derek Abbott}

\address{Centre for Biomedical Engineering (CBME) \\
and Department of Electrical and Electronic Engineering \\
The University of Adelaide, Adelaide, SA 5005, Australia}

\ead{aflitney@eleceng.adelaide.edu.au,
dabbott@eleceng.adelaide.edu.au}

\begin{abstract}
In game theory,
a popular model of a struggle for survival
among three competing agents is
a truel, or three person generalization of a duel.
Adopting the ideas recently developed in quantum game theory,
we present a quantum scheme
for the problems of duels and truels.
In the classical case,
the outcome is sensitive to the precise rules
under which the truel is performed
and can be counter intuitive.
These aspects carry over into our quantum scheme,
but interference amongst the players' strategies can arise,
leading to game equilibria different from the classical case.
\end{abstract}

%\keywords{game theory, quantum games, duels, truels}


\submitto{\JOB}

%\maketitle

\section{Introduction}
The study of quantum games is motivated by
a desire to understand the nature of quantum information~\cite{god}
and the possibility that it may lead to new or improved
algorithms for quantum computers~\cite{lee02}.
Also, in the field of quantum communication, optimal
quantum eavesdropping can be treated as a strategic game with the goal of
extracting maximal information~\cite{brandt98}.
A truel, and its $n$--player generalization the $n$--uel,
may be used to model such a multiplayer struggle.
We present a quantum scheme with strong analogies to this
classic game-theoretic problem.

Quantum game theory is an exciting new area
%~\cite{meyer99,eisert99,eisert00,benjamin00a,benjamin00b,flitney02a}
that models the interactions of agents that
are able to utilize quantum strategies,
that is, have the ability to make quantum manipulations.
The study was initiated by Meyer~\cite{meyer99}
who showed that a quantum player could always beat a classical one
in the simple game of penny flip.
A protocol for two player--two strategy games ($2 \times 2$)
with entanglement
was developed by Eisert and co-workers~\cite{eisert99,eisert00,benjamin00a}
and extended to multi-player games by Benjamin and Hayden~\cite{benjamin00b}.
Many problems have now been considered by quantum game
theory~\cite{marinatto00,iqbal01a,johnson01,flitney02a,piotr01a}
and an experimental realization of quantum prisoners' dilemma
in a liquid nuclear magnetic resonance machine
has been carried out by Du {\em et~al}~\cite{du02}.
For further references and a review of early work in quantum game
theory see Flitney and Abbott~\cite{flitney02b}.

In quantum games, a binary choice of move by a player is encoded by a qubit,
with the computational basis states $|0\rangle$ and $|1\rangle$
corresponding to the classical moves.
Players carry out local unitary operations on their qubit.
The coherence of the system is maintained until
all players have completed their moves.
Then a measurement is carried out on the final state
and the payoffs are obtained from the classical payoff matrix.
By entangling the players' qubits,
the protocol developed by Eisert {\em et al}
produces results different from those obtainable through
mixed classical strategies.
Players can utilize the increased strategic space available through
the use of superpositions,
as well as entanglement between the agents' actions,
to give effects not seen in classical game theory
including new game equilibria.

\section{Classical truel}
\label{sec-classical}
In the classic wild Western duel,
two gunfighters shoot it out
and the winner is the one left standing.
This situation presents few game theoretic difficulties
for the participants:
shoot first and calculate the odds later is always the best strategy!
When this situation is generalized to three or more players
the situation is more complex and
an intelligent use of strategy can be beneficial.
For example,
Alice, Bob and Charles decide to settle their difference with a shoot out,
firing sequentially in alphabetic order.
Consider the case where Alice has a one-third chance of hitting,
Bob two-thirds, and Charles never misses.
Bob and Charles will both target their most dangerous opponent: each other.
Clearly Alice does not want to hit Bob with her first shot since then she
is automatically eliminated by Charles.
Surprisingly, Alice is better off abstaining (or firing in the air)
in the first round.
She then gets the first shot in the resulting duel,
a fact that compensates for her poorer marksmanship.
Precise results for this case are given below.
The paradox of not wanting to fire can been seen most clearly
when all three protagonists are perfect shots.
Alice is advised not to shoot
since after she eliminates one of the others
she automatically becomes the target for the third.
Unless this is the last round,
Bob prefers not to fire as well for the same reason.
If there is an unlimited number of rounds
no one wants to be the first to eliminate an opponent.
The result is a paradoxical stalemate where all survive.

The rules for truels can vary.
Firing can be simultaneous or sequential
in a fixed or random order,
firing into the air can be permitted or not,
and the amount of ammunition can be fixed or unlimited.
In the current discussion we shall make the following assumptions:
\begin{itemize}
\item Each player strictly prefers survival over non-survival.
Without loss of generality we assign a utility of one to a sole
survivor and zero to any eliminated players.
\item Each player prefers survival with the fewest co-players.
That is, the utility of survival in a pair ($u_2$) or in a three-some ($u_3$)
will obey $0 < u_3 \le u_2 \le 1$.
\item Alice, Bob and Charles have marksmanship
(probability of hitting their chosen target)
of $\bar{a} = 1-a,\; \bar{b}=1-b,\; \bar{c}=1-c$, respectively,
independent of their target
and with $0 \le a,b,c < 1$.
There is no probability of hitting a target other than the one chosen.
\item The players get no information on the others' strategies
apart from knowing who has been hit,
and in the quantum model, not even that.
\item Players fire sequentially in alphabetic order
with firing into the air permitted.
\end{itemize}
%where players fire sequentially in reverse order of marksmanship
%and have an unlimited supply of bullets.
%Without loss of generality,
%take Alice to be the worst shot,
%Bob to be an intermediate shot,
%and Charles to be the best marksman.
%That is, let $a < b < c$,
%where $a$, $b$ and $c$ are the probabilities of hitting
%for Alice, Bob and Charles, respectively.
%If the players are allowed to shoot in the air then,
%while all three players live,
%each participant has three (pure) strategies:
%do nothing, i.e., shoot in the air ($N$),
%shoot at the better shot of the two opponents ($B$),
%or shoot at the worst shot of the two opponents ($W$).
%A little thought will tell us that if we end up in a duel
%we would prefer it to be with the poorer shot,
%so strategy $W$ is never optimal.
%At first sight it may seem that playing $N$ is little better.
%It does nothing directly to advance our cause to be the last person standing.
%However, consider the situation where $a = b = c = 1$.

An analysis of classical truel
is provided by Kilgour
for the sequential~\cite{kilgour75}
and the simultaneous case~\cite{kilgour72}.
A non technical discussion is provided by Kilgour and Brams~\cite{kilgour97}.
To get a flavour of some of Kilgour's results we shall consider
the case where the poorest shot fires first and the best last
($\bar{a} < \bar{b} < \bar{c}$)
and ammunition is unlimited.
First the expectation value of Alice's payoff in a duel between Alice and Bob,
with each having $m$ bullets, is calculated:
%(see \fref{fig-duel}):
\begin{equation}
\langle \$_A \rangle_{m} = 1-a \:+\: ab \, \langle \$_A \rangle_{m-1}.
\end{equation}
When $m \rightarrow \infty$,
$\langle \$_{A} \rangle_{m} = \langle \$_{A} \rangle_{m-1}$, hence
\begin{equation}
\langle \$_A \rangle_{\infty} = \frac{1-a}{1-ab}.
\end{equation}
Note that $\langle \$_B \rangle = 1 - \langle \$_A \rangle$.
Using this result, the expectation values
for each player in a truel can be computed.
%(see \fref{fig-truel1} and \fref{fig-truel2}).

There are three important strategic mixes to consider
depending on Alice's strategy.
What ever Alice does,
Bob is advised to shoot at Charles since he is the one that Bob least wants to
fight in a duel,
and Charles, if he survives,
similarly does best by shooting back at Bob.
If Alice fires in the air on her first shot
(or whenever both other players are alive)
Alice is the sole survivor with probability
\begin{equation}
\label{eqn-A0}
P_{A_0} = \frac{1-a}{1-bc} \left[ \frac{1-b}{1-ab} \:+\: \frac{b(1-c)}{1-ac}
		\right].
\end{equation}
If Alice shoots at Bob or Charles
(when she has a choice)
her resulting odds of survival are
\begin{eqnarray}
\label{eqn-AB}
P_{A_B} &= \frac{1-a}{1-abc} \left[ \frac{a(1-b)}{1-ab}
		\:+\: \frac{c(1-a) + ab(1-c)}{1-ac} \right], \nonumber \\
P_{A_C} &= \frac{1-a}{1-abc} \left[ \frac{a(1-b) + b(1-a)}{1-ab}
		\:+\: \frac{ab(1-c)}{1-ac} \right],
\end{eqnarray}
respectively.
From the fact that $b > c$ it follows that $P_{A_C} > P_{A_B}$
so Alice never fires at Bob while Charles is still alive.
To make this example concrete,
consider the case mentioned above: $a=2/3$, $b=1/3$ and $c=0$.
Then $P_{A_0} = 25/63$ which is better than $P_{A_C} = 59/189$
and $P_{A_B} = 50/189$,
meaning that Alice is advised to begin by shooting in the air
and then to shoot at whoever is left standing after the first round.
Surprising, even though Alice is the worst shot,
this strategy will give her a better than one third probability of survival.
Her advantage comes from the fact that she is not targeted until there is
only a pair of players left and she gets the first shot in the resulting duel.
In contrast, Charles has only a $2/9$ chance of emerging as the sole survivor
even though he is a perfect shot!
He has the disadvantage of shooting last
and being the one that the others most want to eliminate.
The results can be sensitive to a minor adjustment of the rules.
For example,
if the number of rounds is fixed,
at some stage Alice may be better served
by helping Bob to eliminate Charles,
particularly if Bob is a poor marksman,
even at the risk of not getting the first shot in a duel with Bob.
However, the paradoxical disadvantage of being the best shot
and the advantage of being the poorest
are common to many truels.

\section{Quantum duels and truels}
\label{sec-quantum}
\subsection{A quantum protocol}
\begin{figure}
\begin{center}
\begin{picture}(160,95)(0,-15)
        \put(0,67){A}
	\put(0,37){B}
	\put(0,7){C}
        \put(10,10){\line(1,0){150}}
        \put(10,40){\line(1,0){68}}
	\put(92,40){\line(1,0){68}}
	\put(10,70){\line(1,0){68}}
	\put(92,70){\line(1,0){68}}
        \put(85,70){\circle*{14}}
	\put(85,40){\circle{14}}
	\put(80,35){\line(1,1){10}}
	\put(90,35){\line(-1,1){10}}
	\put(85,47){\line(0,1){21}}
	\put(70,0){\vector(1,0){40}}
	\put(80,-15){time}
\end{picture}
\end{center}
\caption{Diagram representing the operation of Alice ``firing'' at Bob
in a quantum truel.
The solid lines indicate the flow of information (qubits)
and $\otimes$ is a logical NOT operation
that is only applied if the control qubit (filled circle)
is $|1\rangle$.}

\label{fig-control}
\end{figure}
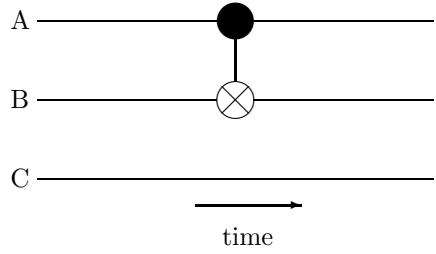

Although the protocol for $2 \times 2$ quantum games has become quite well
established,
the quantization of more a complex game situation is not unique\footnote{
For example, there are three quite different quantizations of the game show
situation
known as the Monty Hall problem~\cite{li01,flitney02c,dariona02}
where a contestant has to guess behind which of three doors a prize lies.}.
We propose the following model of a quantum truel.
Each player has a qubit designating their state,
with the basis states $|0\rangle$ and $|1\rangle$
representing ``dead'' and ``alive,'' respectively.
The combined state of the players is
\begin{equation}
\label{eqn-state}
|\psi\rangle = |a\rangle \otimes |b\rangle \otimes |c\rangle = |abc\rangle,
\end{equation}
with the initial state being $|\psi_i\rangle = |111\rangle$.
In a quantum duel,
the third qubit is omitted.
In a classical truel
the players are located separately.
However, in the quantum case the qubits representing the states
of the players need to be in the one location so that operations can be carried
out on the combined state.
We envisage, for example, a referee applying operators
with the prior instruction of the players.
The analogue of firing at an opponent
will be the attempt to flip the opponent's qubit
using a unitary operator acting on $|\psi\rangle$.
In a duel between Alice and Bob,
the action of Alice ``firing'' at Bob
with a probability of success of $\bar{a} = \sin^2(\theta/2)$
can be represented,
with maximum generality,
by the operator
\begin{eqnarray}
\label{eqn-flip}
\hat{A}_{B} &= \left[e^{-\rmi \alpha} \cos(\theta/2) |11\rangle
		\,+\, \rmi e^{\rmi \beta} \sin(\theta/2) |10\rangle \right] \langle 11|
\nonumber \\
	    & \quad + \left[ e^{\rmi \alpha} \cos(\theta/2) |10\rangle
		\,+\, \rmi e^{-\rmi \beta} \sin(\theta/2) |11\rangle  \right] \langle 10|
\nonumber \\
	    & \quad + |00\rangle \langle 00| + |01\rangle \langle 01|,
\end{eqnarray}
where $\theta \in [0, \pi]$ is fixed 
and $\alpha, \beta \in [-\pi, \pi]$ are arbitrary phase factors.
The last two terms of \eref{eqn-flip}
result from the fact that Alice can do nothing
if her qubit is in the $|0\rangle$ state.
The operator for Bob ``firing'' at Alice, $\hat{B}_{A}$, is obtained
by reversing the roles of the first and second qubits in \eref{eqn-flip}.
For a truel, similar expressions can
be obtained with the third qubit being a spectator.
For example,
\begin{eqnarray}
\label{eqn-flip3}
\hat{A}_{B} &= \sum_{j} \left\{ \left[e^{-\rmi \alpha} \cos(\theta/2) |11j\rangle
		\,+\, \rmi e^{\rmi \beta} \sin(\theta/2) |10j\rangle  \right]
			\langle 11j| \right.  \nonumber \\
	    & \quad + \left. \left[e^{\rmi \alpha} \cos(\theta/2) |10j\rangle
		\,+\, \rmi e^{-\rmi \beta} \sin(\theta/2) |11j\rangle \right]
			\langle 10j| \right\} \nonumber \\
	    & \quad + \sum_{jk} |0jk\rangle \langle 0jk|
\end{eqnarray}
is the operation of Alice ``firing'' at Bob.
That is, Alice carries out a control-rotation of Bob's qubit
with her qubit being the control
(see \fref{fig-control}).
Firing into the air is represented by the identity operator.
For $\alpha$, $\beta$ and $\theta$ we shall use the subscripts
$1$, $2$ and $3$ to refer to Alice, Bob and Charles, respectively.
The operators given flip between the basis states $|0\rangle$ and $|1\rangle$
but do not invert a general superposition.
A general complementing operation in quantum mechanics cannot be achieved
unitarily~\cite{buzek99,pati01,pati02}.
The truel shall be of a fixed number of rounds
with the coherence of the state being maintained
until a measurement is taken on the final state.
Partial decoherence at each step,
where the players obtain some information about the state of the system,
is a possible extension of our scheme.
Expectation values for the payoffs to Alice, Bob and Charles are, respectively,
\begin{eqnarray}
\fl \langle \$_A \rangle &= |\langle 100|\psi_f \rangle|^2
        \,+\, u_{2}( \, |\langle 110|\psi_f \rangle|^2 + |\langle 101|\psi_f \rangle|^2 )
                        \,+\, u_{3} |\langle 111|\psi_f \rangle|^2, \nonumber \\
\fl \langle \$_B \rangle &= |\langle 010|\psi_f \rangle|^2
        \,+\, u_{2}( \, |\langle 110|\psi_f \rangle|^2 + |\langle 011|\psi_f \rangle|^2 )
                        \,+\, u_{3} |\langle 111|\psi_f \rangle|^2, \nonumber \\
\fl \langle \$_C \rangle &= |\langle 001|\psi_f \rangle|^2
        \,+\, u_{2}( \, |\langle 101|\psi_f \rangle|^2 + |\langle 011|\psi_f \rangle|^2 )
                        \,+\, u_{3} |\langle 111|\psi_f \rangle|^2.
\end{eqnarray}
In what follows,
we shall take the utility of surviving in a pair to be $u_2 = 1/2$
and the utility of surviving in a trio to be $u_3 = 1/3$,
so that the combined payoff of any outcome is one.
We shall talk of a player being eliminated
after a certain number of rounds
if there is a probability of one of their qubit being in the $|0\rangle$ state.
As distinct from the classical case,
however, the qubit may subsequently be flipped back to $|1\rangle$,
so in fact the player has not been removed from the game.
To play a quantum duel or truel,
the players list the operators they are going to use in each round
before the game begins.
In the classical case we made the assumption that the players have
no information about the others' strategies except to know who has been hit.
In the quantum case, since a measurement is not taken
until the completion of the final round,
the players do not even have this information.
Thus deciding on the set of operators to use at the start of the game
is no loss of generality.

\subsection{Quantum duels}
\begin{figure}
\begin{center}
\epsfig{file=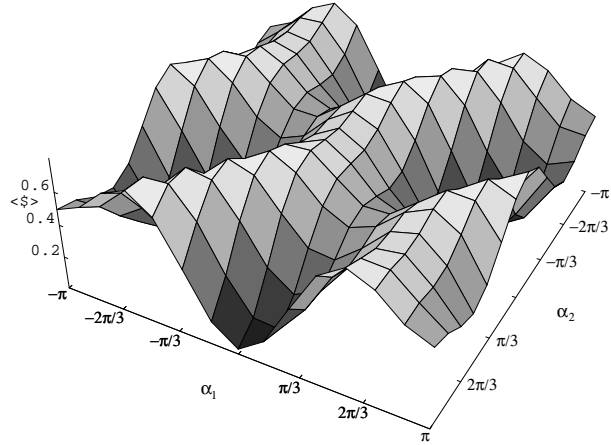,width=8cm}
\end{center}
\caption{The expectation value of Alice's payoff in a two shot quantum duel with Bob,
as a function of $\alpha_1$ and $\alpha_2$,
when the probability of Alice and Bob missing are $a=2/3$ and $b=1/2$, respectively.
The values of $\beta_1$ and $\beta_2$ have no effect.
The $\alpha_i$ and $\beta_i$ are the phase factors from the operator in \eref{eqn-flip}
with the subscript 1 referring to Alice and 2 to Bob.}
\label{fig-phase}
\end{figure}

\begin{figure}
\begin{center}
\epsfig{file=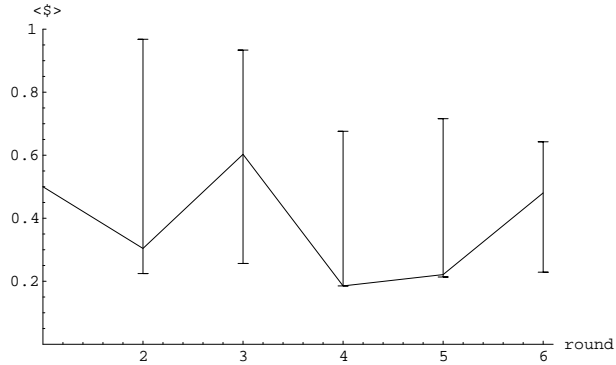,width=8cm}
\end{center}
\caption{The curve shows the expectation value of Alice's payoff
in a repeated quantum duel with $a=2/3$, $b=1/2$
and $\alpha_i = \beta_i = 0$.
The vertical lines indicate the range of possible payoffs
over all values of $\alpha_1$ and $\alpha_2$.
The values of $\beta_1$ and $\beta_2$ have no effect.
For comparison,
a classical duel with the same marksmanship gives Alice and Bob equal chances
(i.e., Alice's payoff is $1/2$).}
\label{fig-duelpayoffs}
\end{figure}

Consider a quantum duel between Alice and Bob.
After $m$ rounds the state of the system will be
\begin{equation}
|\psi_{m}\rangle = (\hat{B}_A \hat{A}_B)^{m} |11\rangle.
\end{equation}
After a single round it is easy to see that a measurement taken at this stage
will not give results any different from the classical duel
with $a=\cos^2(\theta_1/2)$ and $b=\cos^2(\theta_2/2)$.
After two rounds we can begin to see some interference effects:
\begin{eqnarray}
\fl |\langle 01|\psi_2 \rangle|^2
 = (1-b) \left[ ab(1+a) + (1-a)^2
	\:+\: 2 ab \sqrt{a} \, \cos(\alpha_1 + 2 \alpha_2) \right. \nonumber \\
 \left. - 2 a(1-a) \sqrt{b} \, \cos(2 \alpha_1 + \alpha_2)
	\:-\: 2 (1-a) \sqrt{ab} \, \cos(\alpha_1 - \alpha_2) \right], \nonumber \\
\fl |\langle 10|\psi_2 \rangle|^2
 = a(1-a) \left( 1 + b + 2 \sqrt{b} \, \cos(2 \alpha_1 + \alpha_2) \right),
\nonumber \\
\fl |\langle 11|\psi_2 \rangle|^2
 =  1 - |\langle 01|\psi_2 \rangle|^2 - |\langle 10|\psi_2 \rangle|^2.
\end{eqnarray}
The last line is a result of the fact
that there is no possibility of the $|00\rangle$ state.
The expectation value for Alice's payoff can be written as
\begin{equation}
\langle \$_A \rangle = \frac{1}{2}
		(1 + |\langle 10|\psi_2 \rangle|^2 - |\langle 01|\psi_2
\rangle|^2),
\end{equation}
with Bob receiving $1 - \langle \$_A \rangle$.
The value of $a$ and $b$ will determine which of the cosine terms
Alice (or Bob) wishes to maximize.
For example, with $a=2/3$ and $b=1/2$
Alice's payoff is maximized for $\alpha_1 = \pm \pi/3,\, \alpha_2 = \mp 2 \pi/3$
or $\alpha_1 = \pm \pi,\, \alpha_2 = 0$
while Bob's is maximized for $\alpha_1 = 0,\, \alpha_2 = \pm \pi$
or $\alpha_1 = \pm 2 \pi/3,\, \alpha_2 = \mp \pi/3$
(see \fref{fig-phase}).
If the players have discretion over the phase factors,
a maximin strategy for the two round duel
is for the players to select $\alpha_1 = \alpha_2 = \pm \pi/3$
in which case the game is balanced.
The situation for longer duels is more complex.
A classical duel with $a=2/3$ and $b=1/2$ gives
each player a one third chance of eliminating their opponent
in the first round,
with a one-third chance of mutual survival
from which the process repeats itself.
Hence the duel is fair, irrespective of the number of rounds,
Alice's opportunity to fire first compensating for her poorer marksmanship.
\Fref{fig-duelpayoffs} indicates
Alice's payoff for the quantum case as a function of the number of rounds.
The result is affected by the values of $\alpha_1$ and $\alpha_2$
but not by $\beta_1$ and $\beta_2$.

\begin{figure}
\begin{center}
\epsfig{file=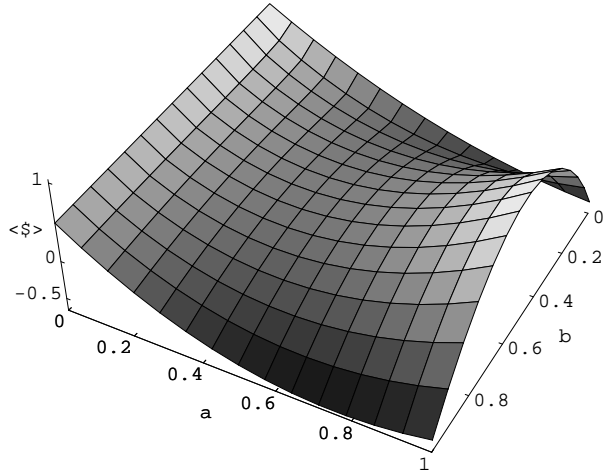,width=8cm}
\end{center}
\caption{In a two shot quantum duel,
the improvement in Alice's expected payoff as a function of $a$ and $b$
(with $\alpha_1 = \alpha_2=0$)
if she chooses to fire in the air on her second shot.
When the value is positive Alice does better by adopting this strategy.}
\label{fig-nosecondshot}
\end{figure}

The fact that a measurement is not taken until the completion
of the game and that the operators are unitary (hence reversible)
means that a $|0\rangle$ state can be unwittingly flipped back to a
$|1\rangle$.
Thus it may be advantageous for one or other player
not to target their opponent.
Consider the situation where Alice fires in the air on her second shot:
\begin{equation}
|\psi_2'\rangle = \hat{B}_A \hat{B}_A \hat{A}_B |11\rangle.
\end{equation}
Then
\begin{eqnarray}
\label{eqn-nosecondshot}
|\langle 01|\psi_2'\rangle|^2 &= 2 ab (1-b) (1 + \sin(2 \alpha_2)), \nonumber \\
|\langle 10|\psi_2'\rangle|^2 &= 1-a.
\end{eqnarray}
If $a$ is sufficiently small
(i.e., Alice has a high probability of flipping Bob's qubit)
then she would prefer this result.
A similar effect holds for Bob if $b$ is small.
Paradoxically, if Alice is a poor shot (approximately $a > 4/5$)
and Bob is intermediate ($b \approx 1/2$)
Alice should refrain from taking a second shot at Bob
(see \fref{fig-nosecondshot}).
\subsection{Quantum truels}
In contrast to the classical case,
players' decisions are not contingent on the success or otherwise of previous
shots.
Since coherence of the system is maintained until the completion of the final
round,
decisions can only be based on the amplitudes of the various states
that the players are able to compute
under different assumptions as to the others' strategies.
The strategies of the other players may be inferred by reasoning
that all players are acting in their self interest.
This idea will guide the following arguments.

In a quantum truel,
interference effects may arise in the first round
if two players choose the same target.
To make the calculations tractable we shall set $\alpha_i = \beta_i = 0;\;
i=1,2,3$
and consider only the case $a > b > c$.
Bob and Charles reason as in the classical case and target each other.
Knowing this, what should Alice do?
If she targets Charles the resulting state after one round is
\begin{equation}
\label{eqn-psi1}
|\psi_1\rangle = (c_1 c_2 - s_1 s_2)(c_3 |111\rangle + s_3 |101\rangle)
			\,+\, (c_1 s_2 + c_2 s_1) |110\rangle,
\end{equation}
where $c_i \equiv \cos(\theta_i/2)$ and $s_i \equiv \sin(\theta_i/2)$.
The probability that Charles survives the combined attentions of Alice and Bob
is $(c_1 c_2 - s_1 s_2)^2$,
compared to the classical case where the probability would be $a b = (c_1 c_2)^2$.
There is much less incentive for Alice to fire in the air since,
unlike the classical case,
%as stated at the beginning of the section,
Bob does not change his strategy (to target Alice)
depending on the results of Alice's operation.
%There is no ``result'' until a measurement is taken.
%Bob can only compute the amplitudes of the various states
%and base his decision accordingly.
If $\theta_1$ and $\theta_2$ are around $\pi/2$ then $c_1 c_2 \approx s_1 s_2$
and both Alice and Bob will like the result of \eref{eqn-psi1}
since Charles has a high probability of being eliminated.

For example, consider the case mentioned in section~\ref{sec-classical}
where $a = (c_1)^2 = 2/3$, $b = (c_2)^2 = 1/3$ and $c = (c_3)^2 = 0$.
If both Alice and Bob target Charles, he is eliminated with certainty
in the first round and consequently his strategy is irrelevant!
If there are sufficient rounds,
Alice would appear to be in difficulties in the resulting duel
since her marksmanship is half that of Bob's.
In a repeated quantum duel where both players continue firing
this is indeed the case.
However, quantum effects come to her rescue
if Alice fires in the air on her third shot.
The expectation value of her payoff after three rounds
is then improved from 0.448 to 0.761.
Indeed, Bob's survival chances are diminished to such an extend
he is advised to fire in the air on the second and subsequent rounds.
We then reach an equilibrium where it is to the disadvantage of both players
to target the other.
Alice emerges with the slightly better prospects
($\langle \$_A\rangle = 0.554$)
since she has had two shots to Bob's one.

Now, compare this to the option of Alice firing in the air in the first round.
With Bob and Charles targeting each other and Charles being a perfect shot,
after the first round the amplitude of states where both survive is zero.
Since Bob fired first and has better then 50\% chance of success,
the $|110\rangle$ state will have
a larger amplitude than the $|101\rangle$ state
so Alice reasons that it is better for her to target Bob in the second round.
Since only one of Bob and Charles can have survived the first round
they both target Alice in the second.
After two rounds the resulting state is
\begin{equation}
\fl |\psi_2\rangle = \frac{1}{\sqrt{27}} (-\sqrt{6} |001\rangle -
\sqrt{8}|010\rangle
			- \sqrt{6} |100\rangle
		- i |011\rangle + i \sqrt{4} |110\rangle + \sqrt{2}
		  |111\rangle).
\end{equation}
Alice calculates (at the beginning of the game)
that if she survives the first two rounds there is a 50\% chance she is the
sole survivor.
If she now targets one of the others in the third round she is more likely to
flip a $|0\rangle$ state to a $|1\rangle$ than the reverse,
hence she fires in the air.
The argument for Bob and Charles to do likewise 
for the same reason is even more compelling.
Hence, even with a large number of rounds,
all players choose to fire in the air after the second round.
The resulting payoffs are
$\langle \$_A \rangle = 52/162$, $\langle \$_B \rangle = 67/162$
and $\langle \$_C \rangle = 43/162$.
Alice clearly prefers to fire at Charles in the first round over this strategy.
It is rare in a quantum truel that Alice will opt to fire in the air
in the first round.
This is in contrast to the classical situation where this is often
the weakest player's best strategy.

In situations where one player is not eliminated with certainty,
an equilibrium where all three players prefer to fire in the air will generally
arise.
Each player reasons that their operation would increase the amplitude
of the $|1\rangle$ state of their target.

\subsection{One- and two-shot truel}
To clarify some of the differences between the classical and quantum truels
consider the simple cases of one- and two-shot truel where Charles is a perfect
shot.
Where Charles is indifferent as to the choice of targets
he uses a fair coin to decide on the target.
In the quantum case, Charles will use this method to select his desired operator
before any operations are carried out on $|\psi_i\rangle$.
For tractability, 
$\alpha_i = \beta_i = 0$ is assumed.

\begin{figure}
\begin{center}
\epsfig{file=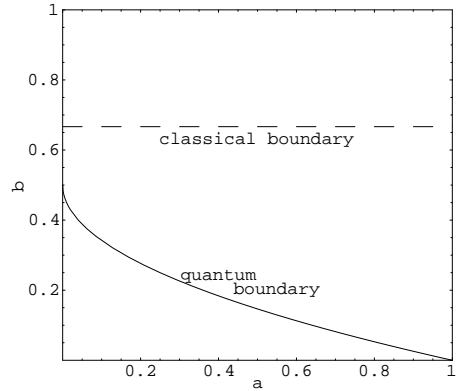,width=6cm}
\end{center}
\caption{In a one-shot truel with $c=0$, Alice's preferred strategy
depending on the values of $a$ and $b$.
Alice fires in the air if $(a,b)$ is below the line
(solid line for the quantum case, dashed line for the classical case)
and at Charles, if above.
The curve is the lower half of $a=(1-2b)^2$.}
\label{fig-onetruel}
\end{figure}

In the one-shot case,
Charles is Bob's only threat so Bob will fire at Charles.
Alice may be targeted by Charles so may wish to help Bob,
particularly if he is a poor shot.
Because of interference,
this strategy is more likely to be preferred in the quantum case.
The regions of the parameter space $(a,b)$
where Alice should select one strategy over the other are indicated in
\fref{fig-onetruel}.
The figure is of interest because it illustrates a case
where going from a classical to a quantum regime
changes a linear boundary in the probability parameter space
into a convex one and such convexity
is being intensely studied as it is the basis of Parrondo's paradox~\cite{harmer02}.

\begin{figure}
\begin{center}
\epsfig{file=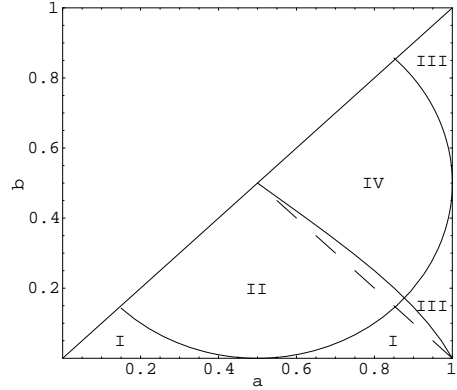,width=6cm}
\end{center}
\caption{In a two-shot truel with $a>b>c=0$, Alice's preferred strategy
depending on the values of $a$ and $b$.
Classical: I and II, fire into the air and then at the survivor of round one;
III and IV, fire at Charles and then at the survivor of round one.
Quantum: I, fire into the air and then at Bob;
II, fire at Charles both times;
III, fire at Charles and then at Bob;
IV, fire into the air and then at Charles.
The boundary between regions I and III or II and IV is the curved line in the
classical case and the dashed line in the quantum case.}
\label{fig-twotruel}
\end{figure}

The situation is more complex in the two-shot case.
When $a>b$,
in the first round Bob and Charles again target each other
while Alice either fires in the air or at Charles.
Since only one of Bob and Charles survive the first round
they both (if alive) target Alice in the second.
In the classical game,
Alice's target in the second round is determined
since she knows whom of Bob or Charles remains.
However, in the quantum case this is unknown and Alice can only base her
decision on maximizing the expectation value of her payoff.
The regions of the parameter space $(a,b)$
where Alice prefers the different strategies are given in
\fref{fig-twotruel}.

\begin{figure}
\begin{center}
\epsfig{file=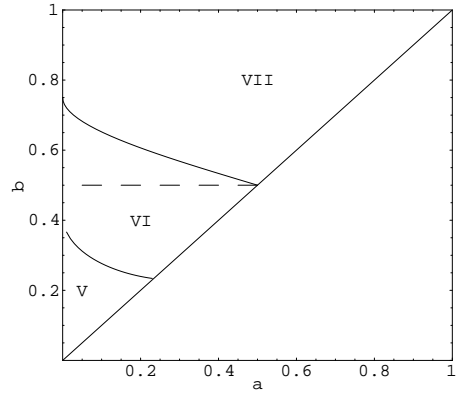,width=6cm}
\end{center}
\caption{In a two-shot truel with $b>a>c=0$, Alice's and Bob's
preferred strategies depending on the values of $a$ and $b$.
Classical: in the first round, Alice fires in the air if $b<1/2$
or at Charles if $b>1/2$.
Quantum: V, Alice fires into the air in round one
and Bob fires at Charles in round two;
VI and VII, Alice fires at Charles in round one
and Bob fires at Alice (VI) or Charles (VII) in round two.}
\label{fig-twotruel2}
\end{figure}

If $b>a$,
Charles will target Alice in the first round
since she is his most dangerous opponent.
Likewise, Bob targets Charles.
In the second round, reasoning as above,
both Alice and Charles (if alive) will target Bob.
In the classical case the only strategic choice is whether
Alice fires at Charles or into the air in the first round.
In the quantum case Bob has a decision to make in the second round
since he does not know for certain who was hit in the first.
\Fref{fig-twotruel2} shows the regions of parameter space
corresponding to Alice's and Bob's optimal choices.

A classical truel where the players do not know which others have
been eliminated may be a fairer comparison to the quantum situation.
This alters the regions corresponding to the players' optimal strategies,
but there are still differences with the quantum truel
as a result of interference in the latter case.


\subsection{Quantum $n$-uels}
A quantum $n$-uel can be obtained by adding qubits to the state $|\psi\rangle$
in \eref{eqn-state}:
\begin{equation}
|\psi\rangle = |q_1\rangle \otimes |q_2\rangle \otimes \ldots \otimes |q_n\rangle,
\end{equation}
where $|q_j\rangle$ is the qubit of player $j$.
The players' operators are the same as \eref{eqn-flip3}
except with additional spectator qubits.
For example,
the first player firing at the second is carried out by
\begin{eqnarray}
\hat{A}_{B} &= \sum_{j_3, \ldots, j_n}
	\left\{ \left[e^{-\rmi \alpha} \cos(\theta/2) |11j_3 \ldots j_n\rangle
                \,+\, \rmi e^{\rmi \beta} \sin(\theta/2) |10j_3 \ldots j_n\rangle  \right]
                        \langle 11j_3 \ldots j_n| \right.  \nonumber \\
            & \quad + \left. \left[e^{\rmi \alpha} \cos(\theta/2) |10j_3 \ldots j_n\rangle
                \,+\, \rmi e^{-\rmi \beta} \sin(\theta/2) |11j_3 \ldots j_n\rangle \right]
                        \langle 10j_3 \ldots j_n| \right\} \nonumber \\
            & \quad + \sum_{j_2, \ldots, j_n}
			|0j_2 \ldots j_n\rangle \langle 0j_2 \dots j_n|,
\end{eqnarray}
where the $j_i$ take the values $0$ or $1$.
%where $\theta \in [0, \pi]$ is fixed
%and $\alpha, \beta \in [-\pi, \pi]$ are arbitrary phase factors.

The features of the quantum $n$-uel are the same
as those of the quantum truel.
Positive and negative interference arising from multiple players
choosing a common target is more likely
and equilibria where it is to the advantage of all (surviving) players
to shoot into the air still arise.

\subsection{Classical-quantum correspondence}
\begin{figure}
\begin{center}
\epsfig{file=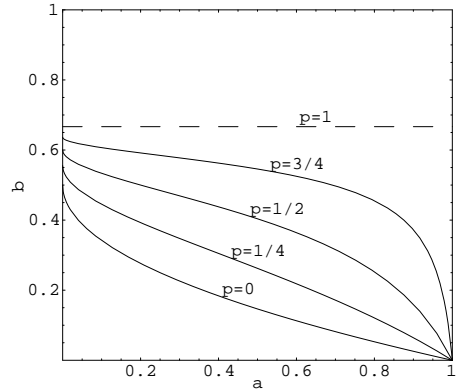,width=6cm}
\end{center}
\caption{In a one shot quantum truel with $c=0$ and with decoherence,
the boundaries for different values of the decoherence probability $p$
below which Alice maximizes her expected payoff by firing into the air
and above which by targeting Charles.
There is a smooth transition from the fully quantum case $(p=0)$
to the classical one $(p=1)$.}
\label{fig-decoherence}
\end{figure}

In the classical case,
players are removed from the game once hit.
Maintained coherence through out the quantum game
weakens the analogy with classical truel
since players can be brought back to ``life,''
that is, have their qubit flipped from $|0\rangle$ to $|1\rangle$.
However, there is still a correspondence.
During the game,
a player can only fire if their qubit is in the $|1\rangle$ state,
and they receive a zero payoff at the end of the game
if their qubit is in the $|0\rangle$ state.
The classical-quantum correspondence can be enhanced by introducing
partial decoherence after each move
and allowing the players to choose their strategy dynamically
depending on the result of previous rounds.
In this case,
the classical situation is reproduced in the limit of full decoherence.
If $\rho = |\psi\rangle \langle \psi|$
is the density operator of the system in state $|\psi\rangle$,
one way of effecting partial decoherence is by
\begin{equation}
\rho \rightarrow (1-p) \rho
	\:+\: p \, {\rm diag}(\rho),
\end{equation}
where $0 \le p \le 1$.
This is equivalent to measuring the state of the system with probability $p$.
When $\rho$ is diagonal,
the next player can select their target based on the measurement results.
Figure~\ref{fig-decoherence} shows the regions of the parameter space $(a,b)$
corresponding to Alice's preferred strategy
in a one shot truel when Charles is a perfect shot
(the situation of figure~\ref{fig-onetruel}).
The boundary between Alice maximizing her expected payoff by firing
into the air and targeting Charles depends on the decoherence
probability $p$.
We then see a smooth transition from quantum case
to the classical one as $p$ goes from zero to one.
Decoherence in quantum games has been considered in a recent publication
by Chen~\etal\cite{chen03}.

\section{Conclusion}
A one round quantum duel is equivalent to the classical game,
but in longer quantum duels the appearance of phase terms in the operators
can greatly affect the expected payoff to the players.
If players have discretion over the value of their phase factors
a maximin choice can in principle be calculated
provided the number of rounds is fixed.
If one player has a restricted choice the other has a large advantage.
The unitary nature of the operators means that the probability of flipping
a ``dead'' state to an ``alive'' state is the same as that for the reverse,
so it can be advantageous for a player to fire in the air rather than target
the opponent,
something that is never true in a classical duel.
Indeed, an equilibrium can be reached where both players forgo targeting
their opponent even if there are further rounds to play.

In a quantum truel, strategies are not contingent on earlier results.
The players' entire strategy
(the list of players to target in different rounds)
can be mapped out in advance based on the expected amplitudes of the various
states
resulting from different strategic choices by the players.
Interference effects arise where a player is targeted by the other two,
and can have dramatic consequences,
either enhancing or diminishing the probability of survival of the targeted
player
compared to the classical case.
As with the case of the quantum duel,
equilibria arise where it is to the disadvantage
of each player to target one of the others.
Such equilibria arise only in special cases in a classical truel.

Introducing decoherence after each move
changes the quantum game.
As the decoherence probability is increased from zero to one (full measurement)
there is a smooth transition from the fully quantum game to the classical one.

\ack
Useful discussions with Arun K. Pati of the Institute of Physics, Orissa, India
are gratefully acknowledged.
This work was supported by GTECH Corporation Australia
with the assistance of the SA Lotteries Commission (Australia).

\end{document}